\documentstyle[12pt,epsf,amssymb]{article}

\def\hybrid{\topmargin -20pt    \oddsidemargin 0pt
        \headheight 0pt \headsep 0pt
        \textwidth 6.25in       
        \textheight 9.5in       
        \marginparwidth .875in
        \parskip 5pt plus 1pt   \jot = 1.5ex}

\hybrid

\def\baselinestretch{1.2}

\catcode`\@=11

\def\marginnote#1{}
\newcount\hour
\newcount\minute
\newtoks\amorpm
\hour=\time\divide\hour by60
\minute=\time{\multiply\hour by60 \global\advance\minute by-\hour}
\edef\standardtime{{\ifnum\hour<12 \global\amorpm={am}%
        \else\global\amorpm={pm}\advance\hour by-12 \fi
        \ifnum\hour=0 \hour=12 \fi
        \number\hour:\ifnum\minute<10 0\fi\number\minute\the\amorpm}}
\edef\militarytime{\number\hour:\ifnum\minute<10 0\fi\number\minute}

\def\draftlabel#1{{\@bsphack\if@filesw {\let\thepage\relax
   \xdef\@gtempa{\write\@auxout{\string
      \newlabel{#1}{{\@currentlabel}{\thepage}}}}}\@gtempa
   \if@nobreak \ifvmode\nobreak\fi\fi\fi\@esphack}
        \gdef\@eqnlabel{#1}}
\def\@eqnlabel{}
\def\@vacuum{}
\def\draftmarginnote#1{\marginpar{\raggedright\scriptsize\tt#1}}

\def\draft{\oddsidemargin -.5truein
        \def\@oddfoot{\sl preliminary draft \hfil
        \rm\thepage\hfil\sl\today\quad\militarytime}
        \let\@evenfoot\@oddfoot \overfullrule 3pt
        \let\label=\draftlabel
        \let\marginnote=\draftmarginnote
   \def\@eqnnum{(\theequation)\rlap{\kern\marginparsep\tt\@eqnlabel}%
\global\let\@eqnlabel\@vacuum}  }

\def\preprint{\twocolumn\sloppy\flushbottom\parindent 2em
        \leftmargini 2em\leftmarginv .5em\leftmarginvi .5em
        \oddsidemargin -.5in    \evensidemargin -.5in
        \columnsep .4in \footheight 0pt
        \textwidth 10.in        \topmargin  -.4in
        \headheight 12pt \topskip .4in
        \textheight 6.9in \footskip 0pt
        \def\@oddhead{\thepage\hfil\addtocounter{page}{1}\thepage}
        \let\@evenhead\@oddhead \def\@oddfoot{} \def\@evenfoot{} }

\def\numberbysection{\@addtoreset{equation}{section}
        \def\theequation{\thesection.\arabic{equation}}}

\def\underline#1{\relax\ifmmode\@@underline#1\else
        $\@@underline{\hbox{#1}}$\relax\fi}

\def\titlepage{\@restonecolfalse\if@twocolumn\@restonecoltrue\onecolumn
     \else \newpage \fi \thispagestyle{empty}\c@page\z@
        \def\thefootnote{\fnsymbol{footnote}} }

\def\endtitlepage{\if@restonecol\twocolumn \else \newpage \fi
        \def\thefootnote{\arabic{footnote}}
        \setcounter{footnote}{0}}  

\catcode`@=12
\relax

\def\figcap{\section*{Figure Captions\markboth
        {FIGURECAPTIONS}{FIGURECAPTIONS}}\list
        {Figure \arabic{enumi}:\hfill}{\settowidth\labelwidth{Figure
999:}
        \leftmargin\labelwidth
        \advance\leftmargin\labelsep\usecounter{enumi}}}
 \relax
\def\tablecap{\section*{Table Captions\markboth
        {TABLECAPTIONS}{TABLECAPTIONS}}\list
        {Table \arabic{enumi}:\hfill}{\settowidth\labelwidth{Table
999:}
        \leftmargin\labelwidth
        \advance\leftmargin\labelsep\usecounter{enumi}}}
 \relax
\def\reflist{\section*{References\markboth
        {REFLIST}{REFLIST}}\list
        {[\arabic{enumi}]\hfill}{\settowidth\labelwidth{[999]}
        \leftmargin\labelwidth
        \advance\leftmargin\labelsep\usecounter{enumi}}}
 \relax
\makeatletter
\newcounter{pubctr}
\def\publist{\@ifnextchar[{\@publist}{\@@publist}}
\def\@publist[#1]{\list
        {[\arabic{pubctr}]\hfill}{\settowidth\labelwidth{[999]}
        \leftmargin\labelwidth
        \advance\leftmargin\labelsep
        \@nmbrlisttrue\def\@listctr{pubctr}
        \setcounter{pubctr}{#1}\addtocounter{pubctr}{-1}}}
\def\@@publist{\list
        {[\arabic{pubctr}]\hfill}{\settowidth\labelwidth{[999]}
        \leftmargin\labelwidth
        \advance\leftmargin\labelsep
        \@nmbrlisttrue\def\@listctr{pubctr}}}
 \relax
\makeatother
\newskip\humongous \humongous=0pt plus 1000pt minus 1000pt

\newif\ifdtup

\relax

\def\be{\begin{equation}}
\def\ee{\end{equation}}
\def\ba{\begin{eqnarray}}
\def\ea{\end{eqnarray}}

\def\del{\partial}

\def\r{\rho}
\def\a{\alpha}

\def\g{\gamma}
\def\G{\Gamma}
\def\d{\delta}

\def\th{\theta}
\def\Th{\Theta}
\def\m{\mu}
\def\n{\nu}

\def\Om{\Omega}
\def\l{\lambda}

\def\S{\Sigma}

\def\cN{{\cal N}}

\def\no{\noindent}

\def\qq{\qquad}

\def\IR{\relax{\rm I\kern-.18em R}}

\def \ha {{1\over 2}}

\def \ov {\over}

\def\IR{\relax{\rm I\kern-.18em R}}
\def\inv{^{\raise.15ex\hbox{${\scriptscriptstyle -}$}\kern-.05em 1}}

\def\tL{{\tilde L}}

\begin{document}

\renewcommand{\theequation}{\arabic{equation}}

\newcommand{\beq}{\begin{equation}}
\newcommand{\eeq}[1]{\label{#1}\end{equation}}
\newcommand{\ber}{\begin{eqnarray}}
\newcommand{\eer}[1]{\label{#1}\end{eqnarray}}
\newcommand{\eqn}[1]{(\ref{#1})}
\begin{titlepage}
\begin{center}

\hfill CERN-TH/99-191\\
\hfill hep--th/9906201\\

\vskip .8in

{\large \bf Wilson loops from multicentre and rotating branes, \\
mass gaps and phase structure in gauge theories}

\vskip 0.6in

{\bf A. Brandhuber }\phantom{x}and\phantom{x} {\bf K. Sfetsos}
\vskip 0.1in
{\em Theory Division, CERN\\
     CH-1211 Geneva 23, Switzerland\\
{\tt brandhu,sfetsos@mail.cern.ch}}\\
\vskip .2in

\end{center}

\vskip .5in

\centerline{\bf Abstract}

\no
Within the AdS/CFT correspondence we use multicentre D3-brane metrics to 
investigate Wilson loops and compute the associated heavy quark--antiquark 
potentials for the strongly coupled $SU(N)$ super-Yang--Mills gauge theory,
when the gauge symmetry is broken by the expectation values of the 
scalar fields. 
For the case of a uniform distribution of D3-branes over a disc, we find that 
there exists a maximum separation beyond which there is no force between 
the quark and the antiquark, i.e. the screening is complete.  
We associate this phenomenon with the possible existence of a mass gap
in the strongly coupled gauge theory.
In the finite-temperature case, when the corresponding 
supergravity solution is a rotating D3-brane solution, 
there is a class of potentials interpolating
between a Coulombic and a confining behaviour.
However, above a certain critical value of the mass parameter, 
the potentials exhibit 
a behaviour characteristic of statistical 
systems undergoing phase transitions. 
The physical path preserves the concavity property of the potential
and minimizes the energy. 
Using the same rotating-brane solutions, we also compute spatial Wilson loops, 
associated with the quark--antiquark 
potential in models of three-dimensional gauge theories
at zero temperature, with similar results.

\no

\vskip 1 cm
\noindent
CERN-TH/99-191\\
June 1999\\
\end{titlepage}
\vfill
\eject

\def\baselinestretch{1.2}
\baselineskip 16 pt
\noindent

\def\tT{{\tilde T}}
\def\tg{{\tilde g}}
\def\tL{{\tilde L}}


\section{Introduction}

The study of $\cN= 4$ $SU(N)$ Yang--Mills (SYM) gauge theory, at large 
$N$ and for large 
\mbox{'t Hooft} coupling constant $R^4=4 \pi g^2_{\rm YM} N$,
is based, in the supergravity approach, on the type-IIB supergravity solution
representing $N$ coincident D3-branes. The metric is
\be
 ds^2 =  H^{-1/2}(-dt^2 + dy_1^2 + dy_2^2 + dy_3^2) + H^{1/2}
(dx_1^2 + \ldots + dx_6^2) \ ,
\label{de3}
\ee
where the harmonic function parametrizing the solution is given by 
\be 
H= 1+ {R^4\a'^2 \ov r^4}\ ,\quad r^2\equiv x_1^2+x_2^2 +\dots + x_6^2\ .
\label{de33}
\ee
The constant string coupling is $e^\Phi = g_s =g^2_{\rm YM}$ and we have
omitted the self-dual five-form.
The field theory correspondence \cite{Maldacena,Gubser,Witten}
is obtained by going to the near-horizon limit
\be
U={r\ov \a'}\ ,\qq U_i = {x_i\ov \a'}\ ,\quad i=1,2,\dots , 6\ ,\qq \a'\to 0\ ,
\label{ftl}
\ee
where the metric factorizes as $AdS_5 \times S^5$ with both factors having
equal radii $R\sqrt{\a'}$.
When the $SU(N)$ gauge symmetry is broken by non-zero expectation values 
(vev's) for the scalar fields, then the appropriate supergravity solution has
a metric given by \eqn{de3}, but with the single-centre 
harmonic function \eqn{de33} replaced by the multicentre one
\be
H = 1+ 4\pi g_s\a'^2 \sum_{i=1}^N {c_i\ov |{\bf x}-{\bf x}_i|^4} \ ,\qq 
\sum_{i=1}^N c_i = N\ ,
\label{hhwp}
\ee
while the dilaton remains constant.

In the AdS/CFT correspondence, isometries of the background are
identified with global symmetries of the gauge theory.
A generic choice of vectors ${\bf x}_i$ completely 
breaks the $SU(4) \simeq SO(6)$ $\cal{R}$-symmetry of the theory. 
However, in this paper we consider
cases where the $\cal{R}$ symmetry group is only partially broken
to $SO(4)$ times a discrete subgroup. This is achieved by placing
the D3-branes in a plane lying in the six-dimensional space transverse
to the D3-branes. It is worth mentioning that in the field theory limit 
\eqn{ftl} there is no supersymmetry enhancement since the (super)conformal
invariance is broken by any non-zero vev's and only ${\cal N}=4$ 
super-Poincar\'e invariance remains.
An important question is whether
there exist supergravity descriptions of the 
same theories with broken gauge group at finite temperature.
The corresponding solutions must be non-extremal versions of 
the backgrounds corresponding to multicentre metrics. 
The construction of 
such solutions is in general not possible since, at non-extremality, the
gravitational attraction of the non-BPS branes 
is no longer balanced by their RR-repulsion.
However, this is possible when we are dealing with rotating D3-brane 
solutions \cite{cvsen,russo,KLT,RS}, 
when the attractive forces are balanced by centrifugal forces.
In this paper we will use  a class of such solutions, which has one rotation 
parameter \cite{russo} that 
breaks the $\cal{R}$-symmetry $SO(6)$ of \eqn{de3} with \eqn{de33}, to an
$SO(4)\times U(1)$ subgroup. In this case the extremal limit of the solution 
describes a uniform distribution of D3-branes over a disc \cite{KLT,Sfe1}.

In this paper we are interested in computing the quark--antiquark 
potential using rotating D3-branes, since multicentre supersymmetric D3-brane
solutions, with a continuous distribution for the branes,
arise when the extremal limit is taken.
In order to calculate the quark--antiquark potential 
we use a method introduced by \cite{rey,malda}. 
According to this prescription the expectation value of a Wilson loop 
of the field theory living on the boundary is given
by the partition function of a fundamental string in the relevant background
and fixed on the boundary to the contour of the Wilson loop that is
to be calculated. 
In principle this amounts to summing over all possible
surfaces of arbitrary genus with the given boundary conditions and all 
quantum fluctuations.
In the supergravity approximation, which is valid for large 't Hooft 
coupling, 
$R^4= 4\pi N g_{YM}^2$, the problem reduces to finding the minimal
area surface that ends on the contour of the Wilson loop on the
boundary. 
The Wilson loop calculated using this procedure deviates
from the usual definition of Wilson loops in that it includes the adjoint
scalar fields (and the fermions). 
This can be understood as follows: in string theory, 
an external quark (or antiquark) is represented as an open string stretched
from the horizon (location of the D3-branes) to the boundary of the 
AdS space.
Since the string ``pulls'' the D3-branes, these are deformed, which 
amounts to turning on scalar fields of the world--volume theory on
the branes. 
Along the Wilson loop the scalars can take non-trivial paths
in all six transverse directions, but in the following we will consider
Wilson loops for which only the scalar corresponding to the radial
coordinate $U$ is active and the others are chosen to be constant.

To be more specific, we have to minimize the Nambu--Goto action of a 
fundamental string in the relevant supergravity background.
In order
to calculate the potential between a quark and an antiquark, we take
a rectangular loop on the boundary with one side of length $L$ in the 
space direction
and one of length ${\cal T}$ in the (Euclidean) time direction, as in field
theory. Since the configuration
is static, the time integration trivially yields a factor of ${\cal T}$
\be
S_{NG} = \frac{{\cal T}}{2 \pi \alpha'} \int dy \sqrt{g(U) \left( \frac{dU}{dy}
\right)^2 + f(U)/R^4}\ ,
\label{wilac}
\ee
where $g(U) = g_{\tau\tau} g_{UU}$ and $f(U) = R^4 g_{\tau\tau} g_{yy}$,
for supersymmetric theories at zero and finite temperature (section 2 and 3). 
In section 4, where we study 
supergravity duals of non-supersymmetric gauge theories, one of the 
spatial coordinates,
say $y_1$, is taken as the Euclidean time and, hence,
$g(U) = g_{yy} g_{UU}$ and $f(U) = R^4 g_{yy}^2$.

The integrals for the length and energy of the Wilson loop, 
respectively, are computed using standard methods developed
in \cite{malda,rey}. They are given by
\be
L \ =\ 2 R^2 \sqrt{f(U_0)} \int_{U_0}^\infty
dU \sqrt{g(U)\ov f(U)(f(U)-f(U_0))} \ ,
\label{le1}
\ee
and
\be
E_{q{\bar q}} \ =\ 
\frac{1}{\pi} \int_{U_0}^\infty dU \left[ {\sqrt{g(U) f(U)\ov 
f(U) - f(U_0)}} - \sqrt{g(U)} \right] - {1\ov \pi}
\int_{U_{\rm min}}^{U_0} dU \sqrt{g(U)} \ ,
\label{en1}
\ee
where $U_0$ denotes the lowest value of $U$ that can be reached by a certain 
string geodesic, and
$U_{\rm min}$ is the minimal $U_0$ allowed by the physics and the geometry
of the particular application. This 
could be the radius of
the disc in the case of the corresponding multicentre metric or
the horizon in the case of non-zero temperature.
Note that we have chosen the arbitrary constant that may be  
added to the energy given by \eqn{en1}, in such a way that the energy of a 
non-interacting pair of quark and antiquark is zero. 
From the form of the multicentre harmonic \eqn{hhwp} it is clear that our
multicentre metrics approach, for large $U$, the metric corresponding to 
$AdS_5\times S^5$ and the same would also be true for the rotating 
brane solutions.
In the conformal case \cite{rey,malda} we have $f(U)=U^4$, $g(U)=1$,
$U_{\rm min}=0$ and we find for the length
\be
L\ =\ {2\sqrt{\pi} \G(3/4)
R^2\ov \G(1/4) U_0}\ ,
\label{sjhd}
\ee
and for the energy 
\ba
E_{q{\bar q}}& =& -{\G(3/4)\ov \sqrt{\pi} \G(1/4)} U_0
\nonumber \\
& =&  - {2 \Gamma(3/4)^2 R^2 \ov \Gamma(1/4)^2}\ {1\ov L}\ .
\label{qqD3}
\ea

One of the issues that we will investigate in detail in section 3 and 4 
is whether the 
concavity condition on the potential of a heavy quark--antiquark pair 
\ba
{d E_{q\bar q}\ov d L} & > & 0\ ,
\label{po1}\\
{d^2 E_{q\bar q}\ov dL^2} & \leq & 0\ ,
\label{po2}
\ea
is obeyed. In physical terms, \eqn{po1} and \eqn{po2} mean that the force 
between the quark and the antiquark is always attractive and a non-increasing 
function of their separation distance. These conditions were proved in 
general in \cite{Bachas}, building on work in \cite{grosse}.
Hence, they should be satisfied by 
the potentials computed using the AdS/CFT correspondence at any order of 
approximation and in particular in the supergravity approximation.
Using \eqn{le1} and \eqn{en1} we find that 
\ba
{d E_{q\bar q}\ov d L} & = & {\sqrt{f(U_0)}\ov 2\pi R^2}\ > \ 0\ ,
\label{pou1}\\
{d^2 E_{q\bar q}\ov dL^2} & = &  {1\ov 4\pi R^2} {f'(U_0)\ov \sqrt{f(U_0)}}
{1\ov L'(U_0)}\ ,
\label{pou2}
\ea
where the prime denotes a derivative with respect to $U_0$.
Hence, the force remains always attractive, except at the point where 
$f(U_0)=0$ and \eqn{po1} is always satisfied. 
However, the concavity condition \eqn{po2} might fail since, even though
in all of our examples $f'(U_0)>0$, there can be 
occasions where $L'(U_0)$ changes sign.\footnote{This, apparently, 
is in conflict with theorem 1 in \cite{Kinar}. Given the explicit
nature of our calculations in sections 3 and 4,
it seems that the validity of this theorem be 
restricted for $E_{q\bar q}< 0$.} 
We will see that such 
behaviour occurs only on non-physical branches of the potential and
was already encountered in previous examples \cite{wilfinTRey,b1}.
We also note that, because of \eqn{pou1}, $L$ and 
$E_{q\bar q}$ reach their extrema at the same value of $U_0$.

The paper is organized as follows: in section 2 we
study Wilson loops in ${\cal N}=4$ SYM with broken
gauge symmetry, using the background for D3-branes
uniformly distributed on a disc. We show that there
is a screening behaviour such that the potential vanishes at 
a finite distance. We argue that this signals
the existence of a mass gap in the theory at strong coupling. 
In section 3
we consider Wilson loops in ${\cal N}=4$ SYM at
finite temperature, using rotating D3-brane solutions
and work out the phase structure in detail.
In section 4 we use the same rotating D3-brane backgrounds
to study non-supersymmetric 
gauge theories in three dimensions; as expected, we find confinement, 
but also a phase transition, depending
on the ratio of the mass to the rotation parameter. 
We present our conclusions and some open problems in section 5.
We have also written an appendix, where we briefly discuss Wilson loops 
obtained using the background for D3-branes distributed uniformly 
on the circumference of a ring.

\section{D3-branes distributed on a disc}

As our first example we consider 
$N$ D3-branes distributed, uniformly in the angular direction,
inside a disc of radius $r_0$ in the  $x_5$--$x_6$ plane.
Their centres are given by \cite{Sfe1}
\ba
&& {\bf x}_{ij} = (0,0,0,0,r_{0j} \cos\phi_i, r_{0j} \sin\phi_i )\ ,
\nonumber \\
&& \phi_i = {2\pi i\ov \sqrt{N}}\ ,
\quad r_{0j}= r_0 \left(j/\sqrt{N}\right)^{1/2}\ ,
\quad  i,j =0,1,\dots,\sqrt{N} -1\ ~, 
\label{vijk}
\ea
and uniformity of the distribution means that we take the constants $c_i$
appearing in \eqn{hhwp} equal to 1.
Since we are interested in the large-$N$ limit, we may take $\sqrt{N}=
{\rm integer}$ without loss of generality. For a number of D3-branes, 
$N\gg 1$ and in the field theory 
limit \eqn{ftl},
where in addition we keep finite the energy of 
strings stretched between D3-branes situated at the centres by 
rescaling $r_0\to \a' r_0$, we may replace the sum by an integral;  
the metric then takes the form \eqn{de3} 
with harmonic given by \cite{Sfe1}
\ba
&&H = \frac{2 R^4/\alpha'^{2}}{\sqrt{(U^2+r_0^2)^2-4 r_0^2 u^2} 
\left( U^2 - r_0^2 + \sqrt{(U^2+r_0^2)^2-4 r_0^2 u^2} \right) } 
\nonumber\\
&& U^2 = U_1^2 + \ldots + U_6^2 \ ,\qq u^2 = U_5^2 + U_6^2 \ .
\label{jew}
\ea
It is easy to see that the above harmonic 
is indeed singular inside a disc of radius $r_0$ in the $x_5$--$x_6$ plane.
This distribution of D3-branes is what is obtained \cite{KLT,Sfe1}
in the extremal limit
of the D3-brane rotating solution with one rotation parameter of \cite{russo}.

\subsection{Trajectory orthogonal to the disc}

There are two simple trajectories we use to investigate the quark--antiquark
potential. 
First consider the trajectory in the hyperplane defined by 
\be 
x_5 = x_6  =  0\ ,\qq u=0\ ,
\label{hs}
\ee
that is along an axis passing through the centre of 
the disc. In that case we have
\be
f(U) = U^2 (U^2 + r_0^2)\ , \qq g(U)=1 
\label{kwp}
\ee
and the minimum value for $U_0$ is $U_{\rm min} = 0$.
The expression for the length as a function of $U_0$ is given by 
(we change the integration variable 
in \eqn{le1} and \eqn{en1} as $\r=U^2$)\footnote{In what follows 
${\bf K}(k)$, ${\bf E}(k)$ and ${\bf \Pi}(n,k)$ 
are complete integrals of the first, second
and third kind, respectively. In this paper we follow the conventions of
\cite{tipologio,BF}.}
\ba
L &=& R^2 U_0 \sqrt{r_0^2 + U_0^2} \int_{U_0^2}^{\infty} \frac{d\rho}
{\rho \sqrt{(\rho + r_0^2)(\rho - U_0^2)(\rho + r_0^2 + U_0^2)}} \nonumber\\
&=& {2 R^2 U_0 k'\ov U_0^2+r_0^2} \left({\bf \Pi}(k'^2,k) 
- {\bf K}(k)\right)\ ,
\label{lle}
\ea
whereas for the quark--antiquark potential as a function of $U_0$, we obtain 
\ba
E_{q{\bar q}} &=& \frac{1}{2 \pi} \int_{U_0^2}^{\infty} d\rho \left[ 
\sqrt{\frac{\rho + r_0^2}{(\rho - U_0^2)(\rho + r_0^2 + U_0^2)}} - 
\frac{1}{\sqrt{\rho}} \right] - \frac{U_0}{\pi}
\nonumber \\
&=& {\sqrt{2 U_0^2+ r_0^2}\ov \pi} \left(k'^2 {\bf K}(k) - {\bf E}(k)\right)\ ,
\label{eenq}
\ea
with $k = \frac{U_0}{\sqrt{r_0^2 + 2 U_0^2}}$ and $k'=\sqrt{1-k^2}$.
For the disc metric we find Coulombic behaviour 
for small $L$, as expected, but an expansion of the integrals for 
small $U_0$ reveals a quite different behaviour for larger values of $L$.
We find that 
\be 
L  = {\pi R^2 \ov r_0} \left( 1 - {U_0\ov r_0}\right) \ +\ {\cal O}(U_0^2)
\label{hf1}
\ee
and 
\be 
E_{q{\bar q}}  =  - {U_0^2\ov 4 r_0} \ +\ {\cal O}(U_0^4)\ .
\label{hf2}
\ee
Hence, we see that there exists a maximum length 
\be 
L_{\rm max} = {\pi R^2\ov r_0}\ ,
\label{maxx}
\ee
after which there is no force to keep the quark and antiquark together.
Solving for $U_0$ in terms of $L$, using \eqn{hf1}, we find that 
\be
E_{q{\bar q}}  =  - {r_0^3\ov 4 } \left({L_{\rm max} - L \ov \pi R^2}\right)^2
\Theta(L_{\rm max}-L) \ +\ {\cal O}(L_{\rm max}-L)^4\ ,
\label{wer}
\ee
where the step function $\Th(L_{\rm max}-L)$ enforces the maximum length 
condition.
The potential goes to zero for a finite separation $L_{\rm max}$,  
which corresponds
to a trajectory that goes all the way down to $U=0$ i.e. touches 
the disc.

The heavy quark--antiquark potential energy as a function of 
the separation is depicted by curve (b) in fig. 1.

\subsection{Trajectory lying on the plane of the disc}

For the trajectory along an axis that lies on the plane of 
the disc and goes through its centre, we have 
\be
x_1 =  x_2 =  x_3 =  x_4  =  0\ ,\qq U = u
\label{jks}
\ee
and
\be
f(U) = (U^2 - r_0^2)^2 \ , \qq g(U)=1 \ ,
\label{rtw}
\ee
whereas $U_{\rm min}=r_0$. The latter condition means that no trajectory
can penetrate the disc.
After we change the integration variable 
as $\r=U^2$ the integral \eqn{le1} for the length becomes
\be
L = R^2 (U_0^2 - r_0^2) \int_{U_0^2}^\infty
\frac{d\rho}{(\rho - r_0^2)\sqrt{\rho (\rho - U_0^2) (\rho + U_0^2
- 2r_0^2)}} \ .
\label{tora}
\ee
Similarly, that of the quark--antiquark potential in \eqn{en1} becomes
\be
E_{q{\bar q}} = \frac{1}{2 \pi} \int_{U_0^2}^\infty
d\rho \left( \frac{\rho - r_0^2}{\sqrt{\rho (\rho - U_0^2) 
(\rho + U_0^2 - 2 r_0^2)}} - \frac{1}{\sqrt{\rho}} \right)
- \frac{U_0-r_0}{\pi}\ .
\label{hwp}
\ee
They can be written in terms of elliptic integrals of
the first and the third kind as
\ba
L = \left \{ \begin{array} {ccc}
 {2 R^2 (U_0^2-r_0^2)\ov r_0^2 U_0}\ \Big({\bf \Pi}(r_0^2/U_0^2,k_1)
- {\bf K}(k_1)\Big)
~~ & {\rm if }~~ & r_0^2 \leq U_0^2 \leq 2 r_0^2  \\ \\
{\sqrt{2} R^2 \ov \sqrt{U_0^2-r_0^2}}\ \Big({\bf \Pi}(\ha,k_2)
- {\bf K}(k_2)\Big)
~~&{\rm if}~~ & U_0^2\geq 2 r_0^2  \\
\end{array}
\right\} ~ ,
\label{comstru}
\ea
and 
\ba
E_{q{\bar q}} = \left \{ \begin{array} {ccc}
 -{U_0\ov 2\pi} (1+k_1){\bf E}\Big({2 \sqrt{k_1}\ov 1+k_1}\Big) +{r_0\ov \pi}
~~ & {\rm if }~~ & r_0^2 \leq U_0^2 \leq 2 r_0^2  \\ \\
{\sqrt{U_0^2- r_0^2}\ov \sqrt{2} \pi}\ \Big({\bf K}(k_2) - 2 {\bf E}(k_2)\Big)
+ {r_0\ov \pi}
~~&{\rm if}~~ & U_0^2\geq 2 r_0^2  \\
\end{array}
\right\} ~ ,
\label{comst1}
\ea
where 
$k_1 = \frac{\sqrt{2 r_0^2 - U_0^2}}{U_0}$ and 
$k_2 = \sqrt{\frac{U_0^2 - 2 r_0^2}{2 U_0^2 - 2 r_0^2}}$. It can be checked 
that $L$ and $E_{q{\bar q}}$, as given by \eqn{comstru} and \eqn{comst1},
are smooth functions at $U_0=\sqrt{2} r_0$.
The behaviour is similar to that in the previous case. We find\footnote{The 
expansion of the complete elliptic integral of the third kind 
${\bf \Pi}(r_0^2/U_0^2,k_1)$ is facilitated if it is first expressed in terms
of (incomplete) integrals of the first and second kind, using eq. 412.01 of
\cite{BF}.}
\be 
L = {\pi R^2 \ov 2 r_0}\ \left( 1- {2 (U_0-r_0)\ov r_0}\left(\ln\left({4 r_0\ov
U_0-r_0}\right) -1\right)\right) \ + \ {\cal O}(U_0-r_0)^2\ ,
\label{lfgr}
\ee
and 
\be
E_{q{\bar q}} = - {(U_0-r_0)^2\ov 2 \pi r_0}\
\left(\ln\left({4 r_0\ov
U_0-r_0}\right) -{3\ov 2}\right) \ + \ {\cal O}(U_0-r_0)^4\ ,
\label{qjr}
\ee
and hence there is a screening behaviour. 
The maximal separation of the quark--antiquark pair is
\be
L_{\rm max} = \frac{\pi R^2}{2 r_0} \ .
\label{gew}
\ee

The heavy quark--antiquark potential energy as a function of 
the separation is depicted by curve (a) in fig. 1.

\subsection{The mass gap} 

We have seen that in both cases there exists a 
maximum length at which the force between the quark and antiquark becomes 
zero, i.e. complete screening!
For a separation of the quark--antiquark pair $L > L_{\rm max}$ we do not 
find a geodesic
connecting them and the solution is given by two straight
strings corresponding to zero potential and zero force between the
quark and the antiquark. The latter state is in fact the physical 
one and prevents a violation of the condition \eqn{po1}.
A screening-type behaviour is typical of
Wilson loops at finite temperature \cite{b1} and also in theories at 
zero temperature, with running
YM coupling constant \cite{KS} (because of non-constant dilaton), 
but it was not expected for a theory at zero temperature with constant 
dilaton. 
From a physical point of view the existence of a maximum length 
$L_{\rm max} \sim {R^2\ov r_0}$ means that one cannot probe at 
arbitrarily small energy scales. This is a 
signal that the gauge theory at strong coupling
exhibits a mass gap $M_{\rm gap} \sim {r_0\ov R^2}$.
In order to investigate this we start with the massless wave equation 
\be
{1\ov \sqrt{G}} \del_\m \sqrt{G} G^{\m\n} \del_\n \Psi = 0\ .
\label{hgf}
\ee
We make the ansatz $\Psi= \phi(U) e^{i k \cdot y} $ and define the 
(mass)$^2$ as  
$M^2=-k^2$ in the spirit of \cite{Witten2}.
In the background \eqn{jew} we obtain the following second order  
differential equation\footnote{The coordinate $U$ appearing in this subsection
is not the same as the one in \eqn{jew}. The two are related by the coordinate 
transformation \eqn{jwoi}, relating \eqn{ruus2} to \eqn{ruu1},
if we analytically continue $r_0^2\to -r_0^2$.}
for $\phi(U)$ 
\be
\del_U U^3 (U^2+r_0^2) \del_U \phi + R^4 M^2 U \phi = 0 \ ,
\label{hjwi}
\ee
whose general solution, for $q\neq 0$,
can be written in terms of hypergeometric functions as
\ba
\phi & =& C_1\left({r_0\ov U}\right)^{1+q}
F\Big(-{1+ q\ov 2},{3-q\ov 2},1-q;-(U/r_0)^2\Big)
\nonumber\\
&&+\  C_2 \left(r_0\ov U\right)^{1-q}
F\Big(-{1-q\ov 2},{3 +q\ov 2},1+q;-(U/r_0)^2\Big)\ ,
\label{wo11}\\
q & \equiv & \sqrt{1-R^4 M^2/r_0^2} \ ,
\nonumber
\ea
where $C_1$, $C_2$ are constant coefficients. The solution that is
valid for $q=0$
can also be written down using \cite{tipologio,tipolo1}, but it will not
be needed for our purposes.
Depending on the value of $M$, 
the parameter $q$ is real or purely imaginary. We will show that only in the 
latter case may we obtain a basis that is 
orthonormalizable in the Dirac sense (with the use of a $\d$-function)
with measure $dU U$.
Indeed, the behaviour of $\eqn{wo11}$ for small $U$ is
\be
\phi \ \simeq \ C_1\left(r_0\ov U\right)^{1+q} 
+ C_2\left(r_0\ov U\right)^{1-q}\ , 
\qq {\rm as}\quad  U\to 0^+\ .
\label{dsh}
\ee
Hence, if $q$ is real, the orthonormalizability condition requires 
that $C_1=0$. On the other hand if $q$ is imaginary, the reality condition
requires that $C_1$ and $C_2$ be complex conjugate to each other. 
Similarly the behaviour of $\eqn{wo11}$ for large $U$ is 
\ba 
 \phi  & \simeq & C_1\ {\G(1-q)\ov \G(3/2 -q/2)^2} 
+ C_2\ {\G(1+q)\ov \G(3/2 +q/2)^2} + {R^8 M^4\ov 32 U^4}
\nonumber \\
&& \times\ \Bigg[C_1\ {\G(1-q)\ov \G(3/2 -q/2)^2} 
\left(2 \ln \left(U\ov r_0\right) + {3\ov 2}- 2 \g 
- {4 r_0^2\ov R^4 M^2} -2\Psi\Big({1-q\ov 2}\Big) \right)
\label{ajq1}\\
&& +\ C_2\ {\G(1+q)\ov \G(3/2 +q/2)^2}
\left(2 \ln \left(U\ov r_0\right) +  {3\ov 2}- 2 \g 
- {4 r_0^2\ov R^4 M^2} -2\Psi\Big({1+q\ov 2}\Big)\right)\Bigg] \ ,
\nonumber\\
&& \hskip 6cm \phantom{dhsgfsfglgdxg} {\rm as}\quad  U\to \infty\ ,
\nonumber
\ea
where $\g$ is Euler's constant and $\Psi(z)={d\ln G(z)\ov dz}$ 
is the so-called psi function.\footnote{In 
finding the asymptotic behaviours in \eqn{dsh} and \eqn{ajq1}, we have
used the fact that $F(a,b,c;0)=1$, the relation 15.3.14 of \cite{tipolo1},
as well as some properties of the psi function.}
It is clear that orthonormalizability 
requires that the constant term in 
the above expansion vanishes, giving
\be
 C_1\ {\G(1-q)\ov \G(3/2 -q/2)^2} 
+ C_2\ {\G(1+q)\ov \G(3/2 +q/2)^2} = 0 \ .
\label{sja21}
\ee
As we have mentioned, real $q$ requires that $C_1=0$ and, because of
\eqn{sja21}, $C_2=0$  as well. Hence, the only possibility to have 
square integrable solutions is to require an imaginary $q$. Indeed, 
\eqn{ajq1} then behaves as $1/U^4$ for large values of $U$.
Furthermore, the condition that $q$ is imaginary leads to 
a mass gap
\be
M_{\rm gap} = {r_0\ov R^2} \ ,
\label{gpww}
\ee   
which is of the order expected from 
the quark--antiquark potential computations.
The mass spectrum above the mass gap is continuous.


\begin{figure}[htb]
\epsfxsize=4in
\bigskip
\centerline{\epsffile{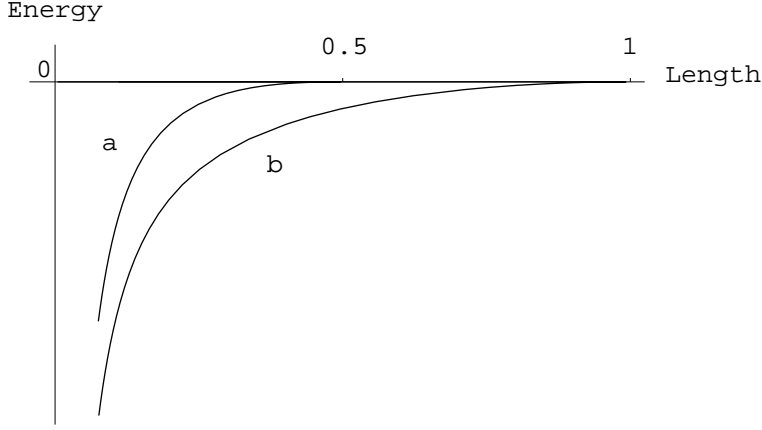}}
\caption{Curve (a) corresponds to the quark--antiquark potential as computed 
using \eqn{comstru} and \eqn{comst1}. 
Curve (b) corresponds to the same 
potential as computed using \eqn{lle} and \eqn{eenq}. 
Lengths and
energies are measured in units of ${\pi R^2\ov r_0}$ and $r_0$, respectively.
Both curves demonstrate that there is a maximum value $L_{\rm max}$,
given by \eqn{gew} and \eqn{maxx}, where the energy becomes zero and the 
screening of charges is complete.
}
\end{figure}


%

\section{Rotating D3-brane solution}

A rotating D3-brane solution of  
type-IIB supergravity was found in \cite{russo} (based on \cite{cvsen}).
For our purposes we pass to the Euclidean regime, where  the time
variable as well as the angular parameter are analytically continued. 
In the field theory limit, the metric is therefore given by 
(the dilaton is constant and we omit the self-dual five-form):
\ba
ds^2 &=& H^{-1/2}\left( f d\tau^2 + dy_1^2 + dy_2^2 +dy_3^2\right) 
+\a'^2 H^{1/2}
\Bigg({dU^2\over f_1}  + (U^2-r_0^2 \cos^2\th)\ d\th^2 
\nonumber\\
&& + (U^2-r_0^2) \sin^2\th\ d\phi^2 
+U^2 \cos^2\th\ d\Omega_3^2-{2 \m^2 r_0\over R^2} 
\sin^2\th d\tau d\phi\Bigg)\ ,
\label{ruu1}
\ea 
with
\ba
H & =&  {R^4/\a'^2 \over U^2(U^2-r_0^2\cos^2\th)}\ , 
\nonumber \\
 f & =&  1-{\m^4\ov U^2(U^2-r_0^2\cos^2\th)}\ ,
\label{defff}\\
f_1& =&  {U^4-r_0^2 U^2 -\m^4\ov 
U^2(U^2-r_0^2\cos^2\th)}\ ,
\nonumber
\ea
where $r_0$ is the ``Euclidean'' angular momentum parameter.
The location of the horizon is given by the positive
root of the equation $U^4-r_0^2 U^2 -\m^4=0$:
\be
U_H^2 = \ha \left( r_0^2+ \sqrt{r_0^4 + 4 \m^4}\right)\ ,
\label{jsab}
\ee
whereas the temperature associated with \eqn{ruu1} is 
\be
T_H = {U_H\ov 2\pi R^2 \m^2} \sqrt{r_0^4+4 \m^4}\ .
\label{sh21}
\ee
Using the expression for the curvature--invariant $R_{\m\n} R^{\m\n}$ 
as given by eq. (3.21) of \cite{russo} 
(analytically continued in the Euclidean regime), we see that the 
metric \eqn{ruu1} has singularities, at $U=0$ and at $U^2=r_0^2 \cos^2\th$, 
which
are hidden inside the horizon at $U=U_H$, provided that $\m>0$.

In this section we are interested in calculating the potential 
between a quark and an
antiquark in a four-dimensional gauge theory at finite 
temperature.\footnote{For a pedagogical review of gauge theories at finite
temperature, see \cite{svet}.}
At finite temperature the time coordinate 
is periodically identified and the rectangular
Wilson loop at zero temperature is wrapped on the time direction.
Therefore the topology of the boundary of the Wilson loop at finite
temperature is that of two circles separated in the space direction
by a distance $L$.
On the supergravity side this means that we have to minimize the
area of a cylindrical string ending on two cycles at the boundary.
Since the configuration is time--independent, formulae
\eqn{wilac}--\eqn{en1} can be used without modification 
to calculate the quark--antiquark potential.

\subsection{Finite temperature and zero rotation}

Before we examine the general case in detail we focus our attention to 
two particular limits, where the metric \eqn{ruu1} simplifies 
considerably and we may compute the integrals in 
\eqn{le1} and \eqn{en1} explicitly.
In the first, the parameter $r_0\ll \m$. 
Then we obtain the non-extremal D3-brane metric in the field theory 
limit given by 
\be
ds^2 = {U^2\ov R^2}\left( \Big(1-{\m^4\ov U^4}\Big) d\tau^2 
+ dy_1^2+ dy_2^2 +dy_3^2
\right) + {R^2\ov U^2}\left( \Big(1-{\m^4\ov U^4}\Big)^{-1} dU^2 + U^2
d\Om_5^2 \right)\ .
\label{jhdg}
\ee
Wilson loops in this background were studied in \cite{wilfinTRey,b1}.
Here we show that various results can be expressed in terms of hypergeometric
functions. We have
\be
f(U) = U^4 -\m^4\ , \qq g(U)=1
\ee
and $U_{\rm min}=\m$. Using \eqn{le1} and \eqn{en1}, we find for the
length:  
\ba
L & =&  2 R^4 \sqrt{U_0^4-\m^4} \int_{U_0}^\infty {dU\ov \sqrt{(U^4 - \m^4)
(U^4-U_0^4)} }
\nonumber \\
& = & \ha B(3/4,1/2)\ {R^2\ov U_0}\ \sqrt{1-\m^4/U_0^4}\ 
F\Big(\ha,{3\ov 4},{5\ov 4};\m^4/U_0^4\Big)
\label{hds}
\ea
and for the energy:
\ba
E_{q{\bar q}} & = & {1\ov \pi} \int_{U_0}^\infty dU\left (\sqrt{{U^4-\m^4}\ov
U^4-U_0^4} -1 \right) -{U_0-\m\ov \pi}
\nonumber \\ 
& = & {B(-1/4,1/2)\ov 4\pi}\ U_0\ 
F\Big(-\ha,-{1\ov 4},{1\ov 4};\m^4/U_0^4\Big) + {\m\ov \pi}\ .
\label{hds1}
\ea
For $U_0\gg \m$ we have the usual Coulombic behaviour, and the potential 
is given by \eqn{qqD3}. As $U_0$ decreases to $U_0\simeq 1.18 \m$ 
the length and energy reach their maximum values
$L_{\rm max}\simeq 0.87 R^2/\m$ 
and $E^{\rm max}_{q{\bar q}}\simeq 0.03 \m$.  
However, before that, for $U_0\simeq 1.52 \m $ 
the energy turns positive and the corresponding
value for the length is $L\simeq 0.75 R^2/\m$.
As $U_0\to \m$ we find that both the length and the energy approach zero 
as\footnote{In the absence of the logarithmic dependence in \eqn{kpw} and 
\eqn{kpw1} the potential energy would have gone to zero as $L^2$ 
for small separations. 
Owing to the presence of logarithm, it is not possible to 
find the analytic expression of the potential energy as a function of the 
length, even if this is small. Nevertheless we easily deduce that 
it goes to zero faster that $L^2$. The authors of \cite{wilfinTRey}
have presented a power--law behaviour of the type $L^{3.6}$ based on
numerical fitting.}
\be
L\ \simeq\ {R^2\ov \m} 
\sqrt{{U_0\ov \m}-1} \left(\ln\left({128\m\ov U_0-\m}\right) -
{\pi\ov 2} \right)\ , 
\label{kpw}
\ee
and
\be 
E_{q{\bar q}}\ \simeq\ {U_0-\m\ov 2\pi}
\left(\ln\left({128 \m\ov U_0-\m}\right) -1 -{\pi\ov 2} \right)\ .
\label{kpw1}
\ee

The plots of the length and potential as functions of $U_0$ and that 
of the potential as a function of the length are given by curve (c) in figs.
2, 3 and 4.

\subsection{The supersymmetric limit and a multicentre metric} 

A second interesting limit is the 
extremal limit corresponding to $r_0\gg \m $, where the solution becomes 
supersymmetric. In this case (strictly speaking exactly for $\m=0$)
the part of the singularity surface at 
$U^2=r_0^2 \cos^2\th$ corresponding to $\th=0$ concides with the 
location of the would--be horizon, which, using \eqn{jsab}, is easily seen 
to be 
at $U_H=r_0$. The nature of this surface, where there will be a distribution 
of D3-branes, becomes transparent 
after the change of variables (valid for the region $U_0\geq r_0$)
\ba
\pmatrix{x_1\cr x_2} &= &  
U \cos\th \sin\psi \pmatrix{\cos\phi_1\cr\sin\phi_1} \ ,
\nonumber\\
\pmatrix{x_3 \cr x_4}&= &  
U \cos\th \cos\psi \pmatrix{\cos\phi_2\cr\sin\phi_2} \ ,
\label{jwoi}\\
\pmatrix{x_5 \cr x_6}& = & 
\sqrt{U^2-r_0^2} \ \sin\th \pmatrix{\cos\phi \cr\sin\phi}\ ,
\nonumber
\ea
where $\psi$, $\phi_1$ and
$\phi_2$ parametrize the line element $d\Om_3^2$ for $S^3$, and we find
\ba
&&ds^2 = H^{-1/2} \left(d\tau^2 + dy_1^2 + dy_2^2 + dy_3^2\right) 
 + H^{1/2} (dx_1^2 +\dots +dx_6^2) \ ,
\nonumber\\
&& H=  {2 R^4/\a'^2\over \sqrt{(U^2-r_0^2)^2+4 r_0^2 u^2}
\left(U^2+r_0^2+\sqrt{(U^2-r_0^2)^2+4 r_0^2 u^2}\right)}\ ,
\label{ruus2}\\
&&U^2=x_1^2+\dots +x_6^2\ ,\qq u^2=x_5^2+x_6^2\ .
\nonumber
\ea
Notice the symbol $U$ in \eqn{jwoi} and \eqn{ruus2} refers to two 
different coordinates. 
The above harmonic function can be obtained from \eqn{jew} if we
analytically continue $r_0\to -i r_0$.
This continuation changes the 
singularity structure of the harmonic function. It no longer represents 
D3-branes uniformly distributed over a disc of radius $r_0$ at the $x_5\!
- \! x_6$ plane, but rather 
D3-branes uniformly distributed over a 3-sphere defined by $x_1^2+x_2^2+x_3^2
+x_4^2=r_0^2$ and $x_5=x_6=0$. 
Notice that the location of the singularity coincides with that of 
the would--be horizon and therefore it is meaningless to talk about 
Hawking temperature  
in this case (note that the expression for $T_{\rm H}$ in \eqn{sh21} 
diverges in the limit $\m\to 0$). 

Before we actually compute the Wilson loops associated with the 
supersymmetric limit, 
let us show that even in that limit there exists a mass gap
and, hence, we expect screening.
As before we start with the massless wave equation \eqn{hgf}, 
with the same ansatz $\Psi= \phi(U) e^{i k \cdot y} $ and 
definition of the (mass)$^2$ as $M^2=-k^2$.
Then we obtain a second order linear differential equation for $\phi(U)$, 
which, after changing variables as 
\be 
\phi = (1-z)^2 Y(z)\ ,\qq z= 1-{2 r_0^2\ov U^2} \ ,\quad |z|\leq 1\ ,
\label{wop3}
\ee
becomes 
\be
(1-z^2) Y'' -2 (1+2 z) Y' + \left({R^4 M^2\ov 4 r_0^2} - 2\right) Y = 0 \ ,
\label{jas}
\ee
which is the Jacobi equation. This has a complete set of normalizable 
solutions in terms of the Jacobi 
polynomials, provided that the mass spectrum is quantized as
\be
M_n^2= {4 r_0^2\ov R^4}\ n (n+1)\ ,\qq n=1,2,\dots \ .
\label{jro}
\ee
In that case $Y(z)\sim P_{n-1}^{(2,0)}(z)$ in the standard notation
of \cite{tipologio}.
Hence, the mass gap must be of the order of the mass associated with the 
lowest eigenvalue $n=1$, which gives a mass gap
$M_{\rm gap} \sim {r_0\ov R^2}$.

In the rest of this subsection we study the potentials arising from two 
different trajectories.

\subsubsection{Case I:}

For the trajectory passing through the centre of the 3-sphere we set 
\be 
x_5  =  x_6 =  0\ ,\qq u=0\ .
\label{saj}
\ee
We have 
\be
f(U) = U^2 (U^2 - r_0^2)\ , \qq g(U) =1
\label{kwp1}
\ee
and $U_{\rm min}=r_0$. 
Then the expression for the length and the potential as functions of $U_0$ are
given by (we change the integration variable 
in \eqn{le1} and \eqn{en1} as $\r=U^2$):
\ba
L &=& R^2 U_0 \sqrt{U_0^2 - r_0^2} \int_{U_0^2}^{\infty} \frac{d\rho}
{\rho \sqrt{(\rho - r_0^2)(\rho - U_0^2)(\rho + U_0^2 - r_0^2)}} \nonumber\\
&=& {2 R^2 U_0 k'\ov U_0^2-r_0^2} \left({\bf \Pi}(k'^2,k) 
- {\bf K}(k)\right)\ ,
\label{lle1}
\ea
and
\ba
E_{q{\bar q}} &=& \frac{1}{2 \pi} \int_{U_0^2}^{\infty} d\rho \left[ 
\sqrt{\frac{\rho - r_0^2}{(\rho - U_0^2)(\rho + U_0^2 - r_0^2)}} - 
\frac{1}{\sqrt{\rho}} \right] - \frac{U_0-r_0}{\pi} 
\nonumber \\
&=& {\sqrt{2 U_0^2- r_0^2}\ov \pi} \left(k'^2 {\bf K}(k) - {\bf E}(k)\right)
+{r_0\ov \pi} \ ,
\label{eenq1}
\ea
where $k = \frac{U_0}{\sqrt{2 U_0^2 -r_0^2}}$ and $k'=\sqrt{1-k^2}$.

For $U_0\gg \m$ we have the usual Coulombic behaviour and the potential 
is given by \eqn{qqD3}. As $U_0$ decreases, for 
$U_0\simeq 1.13 r_0$ 
the length and energy reach their maximum values
$L_{\rm max}= R^2/r_0 $ and $E^{\rm max}_{q{\bar q}}\simeq 0.02 r_0$.
However, before that, for $U_0\simeq 1.38 r_0$,
the energy turns positive and the corresponding
value for the length is $L\simeq 0.88 R^2/r_0$.
As $U_0\to r_0$ we find that 
both the length and the energy approach zero as
\be
L\ \simeq\  {\sqrt{2} R^2\ov r_0} \sqrt{{U_0\ov r_0}-1} 
\left(\ln\left({8 r_0\ov U_0-r_0}\right) -2\right)
\label{jqnr}
\ee
and
\be
E_{q{\bar q}}\  \simeq \ {U_0- r_0\ov 2\pi} 
\left(\ln\left({8 r_0\ov U_0-r_0}\right) -3\right)\ .
\label{jqnr1}
\ee
The plots of the length and potential as a function of $U_0$ and that 
of the potential as a function of the length are given by curve (a) in figs.
2, 3 and 4. We see that the behaviour is similar to the case of finite 
temperature, but zero rotation parameter.

\subsubsection{Case II:}

For the  trajectory with
\be
x_1 = x_2 =  x_3 =  x_4  =  0\ ,\qq U = u\ ,
\label{jks1}
\ee
we have
\be
f(U) = (U^2 + r_0^2)^2\ ,\qq g(U) =1
\label{rtw1}
\ee
and that $U_{\rm min}=0$.
The expressions for the length and the potential as functions 
of $U_0$ are given by (again we change the integration variable 
in \eqn{le1} and \eqn{en1} as $\r=U^2$):
\ba
L & =&  R^2 (U_0^2 + r_0^2) \int_{U_0^2}^\infty
\frac{d\rho}{(\rho + r_0^2)\sqrt{\rho (\rho - U_0^2) (\rho + U_0^2
+ 2r_0^2)}} 
\nonumber\\
& = &  {\sqrt{2} R^2\ov \sqrt{U_0^2+r_0^2}} 
\left({\bf \Pi}\left(\ha,k\right) - {\bf K}(k)\right)
\label{tora1}
\ea
and
\ba
E_{q{\bar q}}& =& \frac{1}{2 \pi} \int_{U_0^2}^\infty
d\rho \left( \frac{\rho + r_0^2}{\sqrt{\rho (\rho - U_0^2) 
(\rho + U_0^2 + 2 r_0^2)}} - \frac{1}{\sqrt{\rho}} \right)
- \frac{U_0}{\pi}
\nonumber\\
& = & {\sqrt{U_0^2+r_0^2}\ov \sqrt{2} \pi} 
\Big({\bf K}(k)- 2 {\bf E}(k)\Big)  \ ,
\label{hwp1}
\ea
where 
$k = \sqrt{\frac{U_0^2 + 2 r_0^2}{2 (U_0^2 + r_0^2)}}$. 
We find that for $U_0\to 0$ 
\be 
L\ \simeq \ { \sqrt{2} R^2\ov r_0} 
\left( \ln\left({4 \sqrt{2} r_0\ov U_0}\right)
- \sqrt{2} \ln(1+\sqrt{2}) \right)
\label{je22}
\ee
and
\be 
E_{q{\bar q}} \ \simeq \ {r_0 \ov \sqrt{2}\pi} 
\left( \ln\left({4 \sqrt{2} r_0\ov U_0}\right) - 2\right)\ .
\label{je23}
\ee
Hence, we find the linear potential 
\be
E_{q{\bar q}}\ \simeq\ \frac{r_0^2}{2 \pi R^2}\ L\ ,\quad {\rm for} \quad 
L\gg {R^2\ov r_0}\ .
\label{conff}
\ee
We note that the potential described by \eqn{tora1} and \eqn{hwp1}
is the same as that found in \cite{minahan} using a two-centre 
metric, even though this is a different supergravity solution than \eqn{ruus2}.

The plot of the potential energy as a function of the quark--antiquark
separation is given by curve (b) in fig. 5. Notice the smooth
interpolation between the Coulombic and confining behaviours.

\subsection{The general case}

We would like to study the quark--antiquark potentials arising at 
general values 
of $\m$ and $r_0$ from trajectories having different, nevertheless constant,
values for the angle $\th$. Since there is  an explicit dependence of the 
metric components on $\th$, consistency requires that the variation of
the Nambu--Goto action with 
respect to $\th$ is zero in order for the 
equations of motion to be obeyed. It is easy to see that this procedure 
allows $\th = 0$ or $\th=\pi/2$. There is no problem setting 
all the other angular variables to constants, 
since the metric components do not explicitly depend on them.

\subsubsection{Trajectory with $\theta = 0$}

In this case
\be
f(U) = U^4 - r_0^2 U^2 - \m^4\ , \qq g(U)=1
\label{jd}
\ee
and the integrals for length and energy, given by 
\eqn{le1} and \eqn{en1} respectively, take the form 
\be
L\ = \ 2 R^2 \sqrt{U_0^4-r_0^2 U_0^2-\m^4} \int_{U_0}^\infty {dU \ov 
\sqrt{(U^4-r_0^2 U^2-\m^4)(U^2-U_0^2)(U^2+U_0^2-r_0^2)}}\ ,
\label{jsa}
\ee
and 
\be
E_{q{\bar q}} \ = \ {1\ov \pi}
\int_{U_0}^\infty dU \left[\sqrt{U^4-r_0^2 U^2 -\m^4\ov
(U^2-U_0^2)(U^2+U_0^2-r_0^2)} -1 \right] -{U_0-U_H\ov \pi}\ ,
\label{dqp}
\ee
with $U_0 \geq U_{\rm min}=U_H$, 
where the location of the horizon is given in \eqn{jsab}.
For large $U_0$ the potential is, as usual, Coulombic, but there
exists a maximal separation, 
as is depicted in fig. 2 for various values of $\m$.
At the values of $U_0$ where the separation is maximal, also the
energy is maximal and always positive (fig. 3).
This means that there is a screening behaviour because, once the
potential turns positive, a configuration of two separate strings,
each bounded by a circle in the Euclidean time direction, 
is energetically favoured and corresponds to a vanishing force
between the charges. As $U_0\to U_H$ we 
find that both the length and the energy approach zero as
\be
L\ \simeq\ \sqrt{2} R^2 (r_0^4+4 \m^4)^{-1/4} \sqrt{{U_0\ov U_H}-1}
\ \ln\left({U_H\ov U_0-U_H}\right)
\label{kpw11}
\ee
and
\be 
E_{q{\bar q}}\ \simeq\ {U_0-U_H\ov 2 \pi}
\ln\left({U_H\ov U_0-U_H}\right) \ .
\label{kp11w1}
\ee
The second branch of the potential, starting at the point of maximal
separation and going to zero distance and zero energy, is unphysical
for three reasons. First, since the energy on this branch is always 
positive the
first branch is preferred, as it it minimizes the energy. Second, it
violates the concavity condition \eqn{po2} (see also \eqn{eqr1} below).
Third, when $U_0 \to U_H$ the circumference of the string in the
Euclidean direction becomes arbitrarily small 
in the region close to the horizon. 
Since the loop is contractible, the string can split into two strings, 
which have the shape of discs bounded by a circle. Therefore,
the potential vanishes, which 
explains the screening when the two quarks are separated
beyond the maximal distance.

\begin{figure}[ht]
\epsfxsize=4in
\bigskip
\centerline{\epsffile{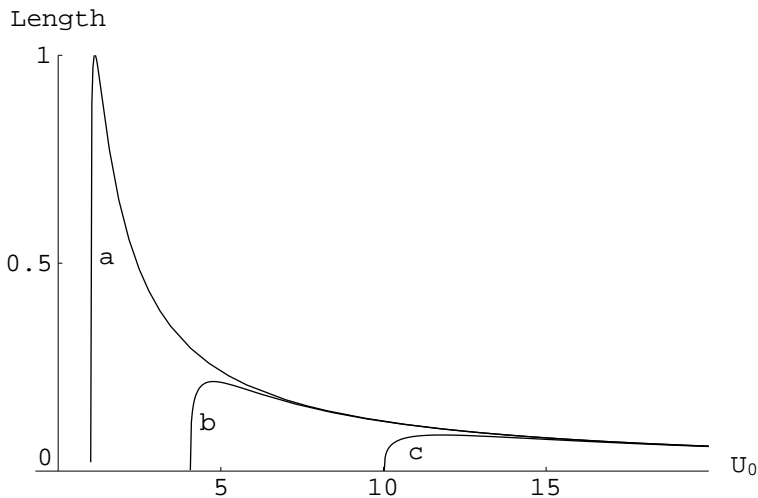}}
\caption{Curves (a), (b) and (c) correspond to the distance
between a quark and an antiquark as a function of $U_0$,
as computed using \eqn{jsa} for three different values of 
$\m=0$, $3$ and $10$, respectively. Lengths and
energies are measured in units of ${R^2\ov r_0}$ and $r_0$, respectively.
All three curves approach $L=0$ as $U_0\to U_H$ according to \eqn{kpw11}.
For large $U_0$ they unify according to \eqn{sjhd} irrespectively
of the value of $\m$. 
Curve (a) can be obtained using \eqn{lle1}. Curve (c) is approximately 
what we would obtain using \eqn{hds}.
}
\end{figure}

\begin{figure}[!ht]
\epsfxsize=4in
\bigskip
\centerline{\epsffile{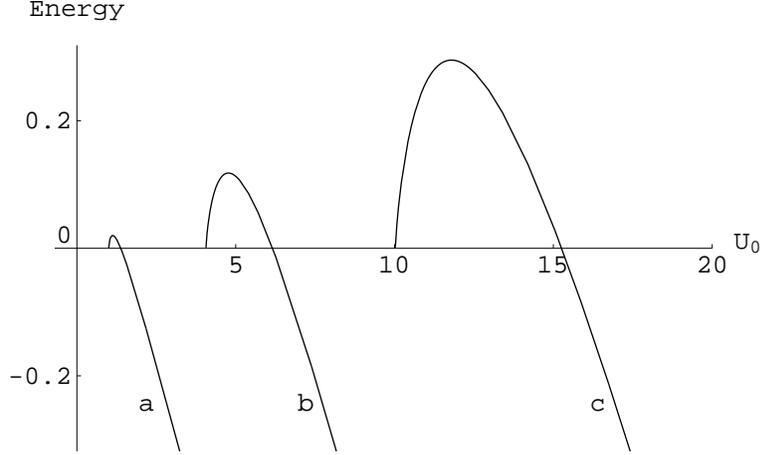}}
\caption{Curves (a), (b) and (c) correspond to the quark--antiquark potential 
as a function of $U_0$ as computed using \eqn{dqp} for three different 
values of 
$\m=0$, $3$ and $10$, respectively. Energies are measured in units of $r_0$.
All three curves approach $U=0$ as $U_0\to U_H$ according to \eqn{kp11w1}. 
For large $U_0$ the curves become parallel as they follow \eqn{qqD3} with
the energy shifted in its curve by ${U_H\ov \pi}$. 
Curve (a) can be obtained using \eqn{eenq1}. Curve (c) is approximately 
what we would obtain using \eqn{hds1}.
}
\end{figure}

\begin{figure}[!ht]
\epsfxsize=4in
\bigskip
\centerline{\epsffile{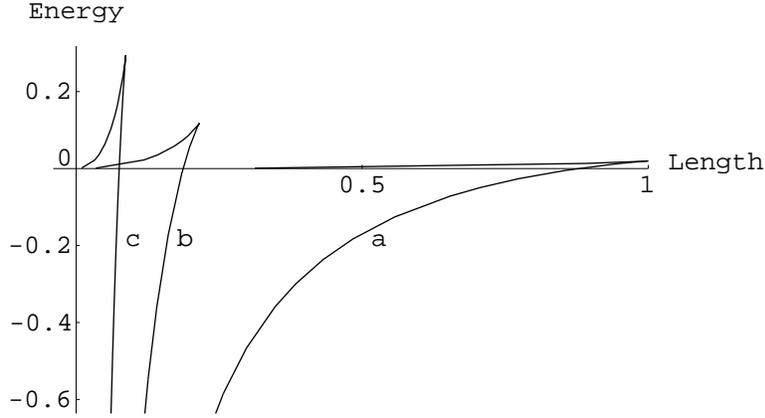}}
\caption{Curves (a), (b) and (c) correspond to the quark--antiquark potential 
as computed using \eqn{jsa} and \eqn{dqp} for the three different values of 
$\m=0$, $3$ and $10$, respectively. Lengths and
energies are measured in units of ${R^2\ov r_0}$ and $r_0$, respectively.
For small separations $L\to 0$, all three curves approach the 
Coulombic law. 
Curve (a) can be 
obtained using \eqn{lle1} and \eqn{eenq1}. Curve (c) is approximately 
what we would obtain using \eqn{hds} 
and \eqn{hds1}.
}
\end{figure}

\newpage

\subsubsection{$\theta = {\pi\ov 2}$}

In this case we have 
\be
f(U) = U^4 - \m^4 \ ,\qq g(U) = (U^4 - \m^4)/(U^4 - r_0^2U^2 - \m^4)\ .
\ee
The integrals for length and the potential energy are given by 
\be 
L \ = \ 2 R^2 \sqrt{U_0^4-\m^4} \int_{U_0}^\infty {dU\ov 
\sqrt{(U^4-r_0^2 U^2-\m^4)(U^4-U_0^4)} }
\label{jk1d}
\ee
and
\ba 
E_{q{\bar q}}& =& {1\ov \pi} \int_{U_0}^\infty dU\left[{U^4-\m^4\ov
\sqrt{(U^4-r_0^2 U^2-\m^4)(U^4-U_0^2)}} 
-\sqrt{U^4-\m^4\ov U^4-r_0^2 U^2 -\m^4} \right] 
\nonumber \\
&& - \ {1\ov \pi} 
\int_{U_H}^{U_0} dU \sqrt{U^4-\m^4\ov U^4-r_0^2 U^2 -\m^4}\ ,
\label{jewh}
\ea
where $U_H$ is given by \eqn{jsab}.\footnote{For $\m=0$ both \eqn{jk1d} 
and \eqn{jewh} tend to the integrals in \eqn{tora1} and \eqn{hwp1} after 
we change variables $U^2\to U^2+r_0^2 $ and rename $U_0^2\to U_0^2+r_0^2 $.}
As usual, the dependence of the potential energy for small separations $L$
of the quark--antiquark (corresponding to large $U_0$) is Coulombic.
For $U_0\to U_H$ we have
\be
L\ \simeq \  {R^2 r_0\ov \sqrt{2} U_H} (r_0^4 + 4 \m^4)^{-1/4} 
\ln\left({U_H\ov U_0-U_H }\right)\ ,
\label{jsd}
\ee
and
\be
E_{q{\bar q}}\ \simeq \  {r_0^2 \ov 2 \sqrt{2} \pi} (r_0^4 + 4 \m^4)^{-1/4} 
\ln\left({U_H\ov U_0-U_H}\right)\ .
\label{j1sd1}
\ee
Eliminating $U_0$ we find a linear confining behaviour 
\be
E_{q{\bar q}}\ \simeq\ \frac{r_0 U_H}{2 \pi R^2}\ L\ ,\quad {\rm for} \quad 
L\gg {R^2 r_0\ov U_H} (r_0^4+4 \m^4)^{-1/4}\ .
\label{conff4}
\ee
However, for intermediate values of the separation,  
the behaviour of the system depends crucially on the value of the
ratio $\l={\m\ov r_0}$. 
There is a critical value $\l_{\rm cr}\simeq 2.85$ 
such that for $\l<\l_{\rm cr}$ the behaviour is
qualitatively the same as in the case $\m=0$ that we have already 
examined (see curve (a) in fig. 5).

\begin{figure}[!ht]
\epsfxsize=4in
\bigskip
\centerline{\epsffile{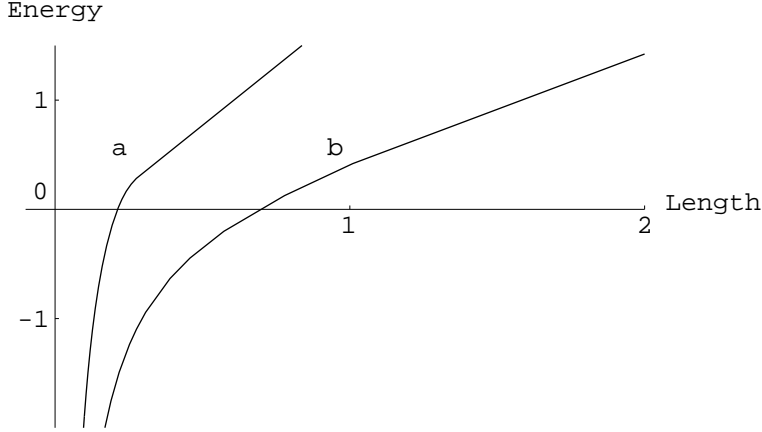}}
\caption{Curves (a), (b) correspond to the quark--antiquark potential 
as computed using \eqn{jk1d} and \eqn{jewh} for the two different values of 
$\m=2$ and $0$, respectively, corresponding to $\l<\l_{\rm cr}\simeq 2.85$. 
Lengths and
energies are measured in units of ${R^2\ov r_0}$ and $r_0$, respectively.
For small separation $L\to 0$, both curves approach the 
Coulombic law. For large separation we obtain a linear behaviour.
Curve (b) can also be 
obtained using \eqn{tora1} and \eqn{hwp1}. 
}
\end{figure}

However, for $\l>\l_{\rm cr}$ the behaviour is different and resembles the 
situation occurring in first order phase transitions in statistical 
systems.\footnote{In terms of thermodynamic quantities, 
our $L$, $E_{q\bar q}$ and $U_0$
correspond to pressure, Gibbs potential and volume, respectively.
At $\l=\l_{\rm cr}$ the first order phase transition
degenerates to a second order one. Then $U_0$ plays the r\^ole of an 
order parameter. For details on the related 
statistical and thermodynamical aspects, see, for instance, \cite{Callen}.
}
We describe the situation below and we also refer to figs. 6 and 7
for graphical details of the length as a function of $U_0$ and 
of the potential energy as a function of the length. In particular,
there are three different branches corresponding to the segments ABCDEF, 
FJKLM and MNOPQRS.
For large $U_0$, equivalently small $L$, 
we have the usual Coulombic behaviour given by \eqn{qqD3} (point A in figs.
6 and 7). 
As we lower $U_0$, we see that for all points until point B there is a unique 
value for $U_0$, that is a unique trajectory,
corresponding to each value for the length.
At point B two different values of $U_0$ correspond to the same value for the
length (the other point is M) and after that, up to point D, there are 
three different $U_0$'s for each value of the length (for instance, C, N
and L). Note also that in the entire path ABCD the energy is the smallest 
compared with the points on the other two branches that have the same length.
When the unique point D is reached, the energy-surface intersects itself 
and after that the energy minimum comes from the other branch of the curve. 
Hence, all three points Q, E and J have the same length, but the physical
state corresponds to point Q since this has the smallest energy of the
three. Therefore, we conclude that the physical path is not the one 
joining the points ADFKMOS but rather ABCDOQRS, where the energy is always a 
minimum for a given value of the length.
Hence, the entire branch FJKLM
connecting the two extrema F and M is unphysical. There is an additional 
reason why this branch is not part of a physical path, namely that 
the concavity condition for the potential \eqn{po2} is violated there. 
This is easily seen by first finding the behaviour close to the two extrema
marked by F and M. Using \eqn{pou1} and \eqn{pou2} we find that 
close to a maximum F 
\be
E_{q\bar q} - E_{q\bar q}^{\rm max}\ \simeq\ {\sqrt{f(U_0^{\rm max})}\ov 
2 \pi R^2}\  (L_{\rm max}-L ) \left[ -1\mp 
D_1 (L_{\rm max}- L)^{1/2}\right] \ ,
\label{eqr1}
\ee
where the minus (plus) sign corresponds to the branch BCDEF (FJKLM)
and $D_1$ is a positive constant.
Close to the minimum M, the corresponding expansion is 
\be
E_{q\bar q} - E_{q\bar q}^{\rm min}\ \simeq\ {\sqrt{f(U_0^{\rm min})}\ov 
2 \pi R^2}\  (L-L_{\rm min} ) \left[ 1\mp 
D_2 (L-L_{\rm min})^{1/2}\right] \ ,
\label{eqr2}
\ee
where now the minus (plus) sign corresponds to the branch MNOQR (FJKLM)
and $D_2$ is another positive constant.
We see that the concavity condition \eqn{pou2} is violated in the branch 
FJKLM. In contrast, in the entire physical path ABCDOQRS this condition
is preserved.
Note also that exactly at the critical value $\l=\l_{\rm cr}\simeq 2.85$ 
the first order degenerates to a second order phase transition.
Then using the fact that not only the first but also the second derivative
of the length vanishes at some critical value $U_0^{\rm cr}$ we find that 
\be
E_{q\bar q} - E_{q\bar q}^{\rm cr} 
\ \simeq\ {\sqrt{f(U_0^{\rm cr})}\ov 2 \pi R^2} 
(L-L_{\rm cr} ) \left(1-
D_3 |L-L_{\rm cr}|^{1/3}\right)\ ,
\label{eqr3}
\ee
for some constant $D_3$. Therefore, using \eqn{pou1}, we find that 
$L-L_{\rm cr}\sim (U_0-U_0^{\rm cr})^3$.
It can also be shown that $U_0-U_0^{\rm cr}\sim
\l-\l_{\rm cr}$ for $\l>\l_{\rm cr}$ and of course it vanishes, as usual
for an order parameter, for $\l<\l_{\rm cr}$.
Hence, the corresponding critical exponents take the classical values 3 and 1.

Finally, we want to comment on the confining behaviour for large
quark--antiquark separation, present for all values of $\mu$ and $r_0$,
which is somewhat unexpected. Once the potential becomes positive, 
a configuration of two separate worldsheets is energetically
preferred. The breaking of the tube--like worldsheet connecting the
quark--antiquark pair is possible, if the proper length of the circle 
$\sqrt{g_{\tau\tau}}/T_H\sim  \sqrt{\a'}$ as
the string approaches the horizon. Using our metric \eqn{ruu1} we find that 
a breaking of the string is
likely when either $\mu \gtrsim \sqrt{R} r_0$ or $r_0 \gtrsim \sqrt{R} \mu$ 
and in these
cases one has screening as in the case of $\theta=0$. On the other
hand, for $r_0 \sim \mu$, the circumference $\sim \sqrt{\a'} R$ is quite
large and the worldsheets on the confining branch of the potential
are stable, although its energy is larger than that of two
disconnected worldsheets.
The answer to this problem might be related to the discussion in 
\cite{minahan}, 
where an (unexpected) linear potential was found to be unstable 
because of extra QCD states.

\begin{figure}[!ht]
\epsfxsize=4in
\bigskip
\centerline{\epsffile{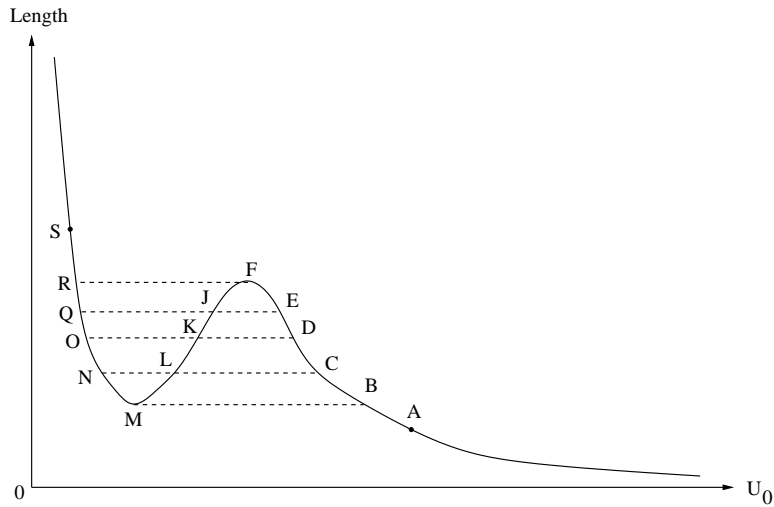}}
\caption{The quark--antiquark separation,   
as computed using \eqn{jk1d} for $\l>\l_{\rm cr}\simeq 2.85$. 
}
\end{figure}

\begin{figure}[!ht]
\epsfxsize=4in
\bigskip
\centerline{\epsffile{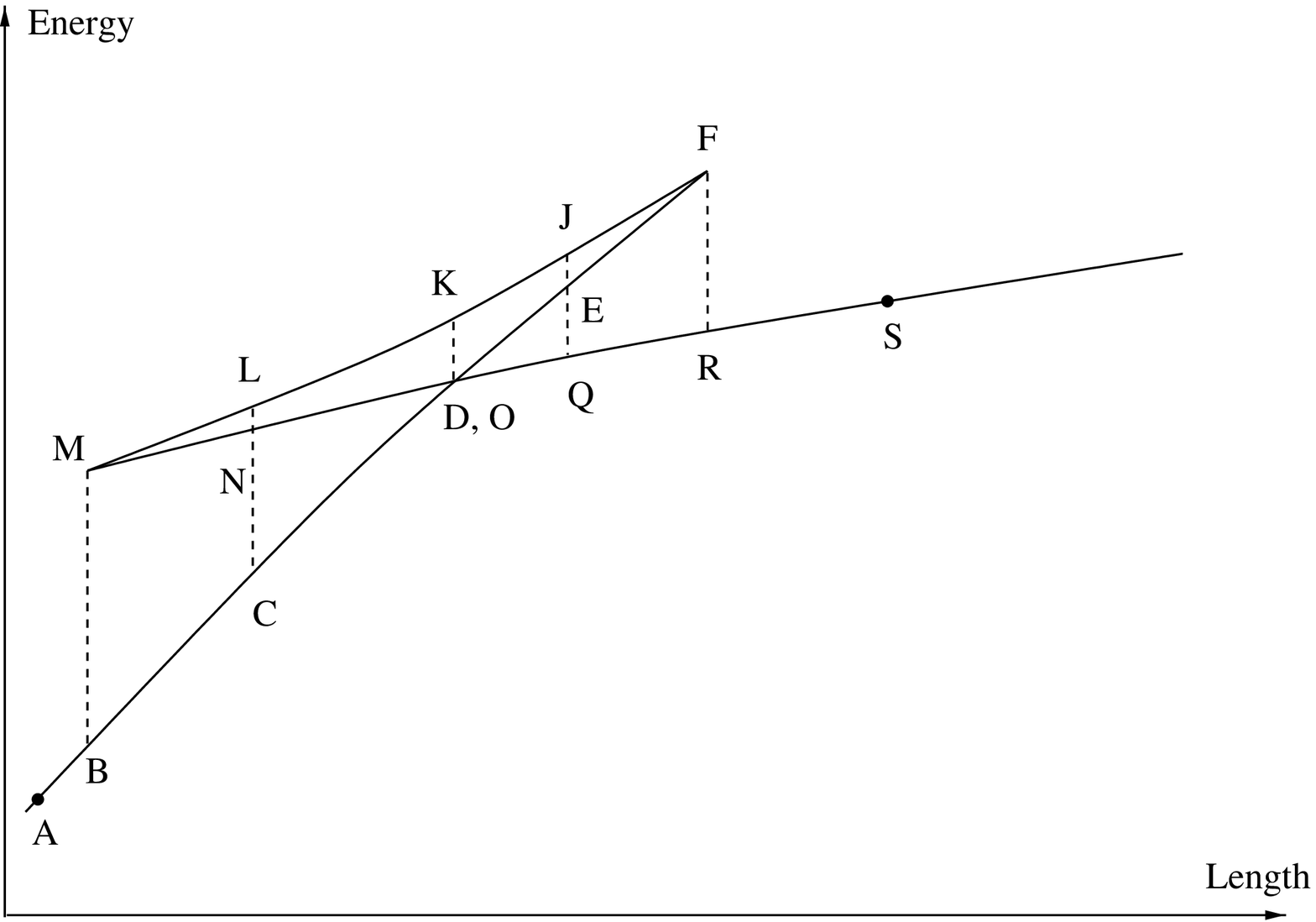}}
\caption{The quark--antiquark potential as a function of the 
separation, 
as computed using \eqn{jk1d} and \eqn{jewh} for 
$\l>\l_{\rm cr}\simeq 2.85$. 
}
\end{figure}

\newpage


\section{Rotating D3-brane for YM in three dimensions}

Models of YM theories in 2+1 dimensions at zero temperature 
can be constructed, using non-extremal D3-branes, by compactifying
the Euclidean time direction on a circle of circumference
$\beta = 1/T_{\rm H}$ \cite{Witten2,COOT98}. 
An extension of this model, where there is a clear decoupling of the
Kaluza--Klein modes associated with the Euclidean time compared with the 
modes that have vanishing Kaluza--Klein charge, is based on rotating 
D3-branes \cite{russo,csaki,RS}. 
The space--space Wilson loop for the zero-rotation case has been computed in 
\cite{b2}.

As before, setting all the angles to constants in searching for 
trajectories is consistent with the equations of motion
only in the cases $\th=0$ or $\th = {\pi\ov 2}$.

\subsection{$\theta = 0$}

In this case we have
\be
f(U) = U^2(U^2 - r_0^2 )\ , \qq 
g(U)=\frac{U^2(U^2-r_0^2)}{U^4 - r_0^2 U^2 - \m^4}\ .
\label{jdconf}
\ee
Then the integrals for the length and the energy, 
given by \eqn{le1} and \eqn{en1} respectively, take the form 
\be
L\ = \ 2 R^2 \sqrt{U_0^2(U_0^2-r_0^2)} \int_{U_0}^\infty {dU \ov 
\sqrt{(U^4-r_0^2 U^2-\m^4)(U^2-U_0^2)(U^2+U_0^2-r_0^2)}}\ ,
\label{jsaconf}
\ee
and 
\ba
E_{q{\bar q}} &=&  {1\ov \pi}
\int_{U_0}^\infty dU \left[ \frac{U^2(U^2-r_0^2)}
{\sqrt{(U^4-r_0^2 U^2-\m^4)(U^2-U_0^2)(U^2+U_0^2-r_0^2)}} -
\sqrt{U^2(U^2 - r_0^2) \ov U^4 - r_0^2 U^2 - \m^4} \right]  
\nonumber \\
&& - \frac{1}{\pi} \int_{U_H}^{U_0} dU 
\sqrt{U^2(U^2 - r_0^2) \ov U^4 - r_0^2 U^2 - \m^4} \ ,
\label{dqpconf}
\ea
with $U_H$ as given by \eqn{jsab}.
For small distances we have, as usual,
a Coulombic behaviour for the potential, 
whereas we find, as $U_0\to U_H$, a linear behaviour since
\be
L\ \simeq\  R^2 \frac{\m^2}{U_H} (r_0^4+4 \m^4)^{-1/2}  
\ln\left({U_H \ov U_0-U_H}\right)
\label{kpw11conf}
\ee
and
\ba 
E_{q{\bar q}} & \simeq & \frac{\m^4}{2\pi U_H} (r_0^4+4 \m^4)^{-1/2}
\ln\left({U_H\ov U_0-U_H}\right) \nonumber \\
& \simeq & \frac{\m^2}{2\pi R^2} L \ ,\qq {\rm for}\quad  
L\gg {R^2 \m^2\ov U_H} (r_0^4+4 \m^4)^{-1/2}  \ .
\label{kp11w1conf}
\ea
Similar to what we have already encountered in section 3, 
there exists a critical value 
$\l_{\rm cr}\simeq 0.46$ 
for the ratio $\l={\m\ov r_0}$,
such that for $\l>\l_{\rm cr}$ the behaviour goes smoothly from Coulombic
to confining.
However, for $\l=\l_{\rm cr}$ there is a second order phase transition 
and  for $\l<\l_{\rm cr}$ the behaviour differs and resembles the 
one depicted in figs. 6 and 7.

It is important to note here that this is not a transition from
confinement to Coulombic potential in a three-dimensional theory, 
but in a four-dimensional theory on a circle. Only for large separation
of the external charges is the theory effectively three-dimensional
and do we observe confinement. But for small distances the string can
probe the extra dimension and the Coulomb behaviour is that of the
compactified four-dimensional theory.

\subsection{$\theta = {\pi\ov 2}$}

In this case we have 
\be
f(U) = U^4 \ ,\qq g(U) = \frac{U^4}{U^4 - r_0^2U^2 - \m^4}\ .
\ee
Then the integrals for the length and the potential energy are given by
\be 
L \ = \ 2 R^2 U_0^2 \int_{U_0}^\infty {dU\ov 
\sqrt{(U^4-r_0^2 U^2-\m^4)(U^4-U_0^4)} }
\label{jk1dconf}
\ee
and
\ba 
E_{q{\bar q}}& =& {1\ov \pi} \int_{U_0}^\infty dU\left[{U^4 \ov
\sqrt{(U^4-r_0^2 U^2-\m^4)(U^4-U_0^2)}} 
-{U^2 \ov \sqrt{U^4-r_0^2 U^2 -\m^4}} \right] 
\nonumber \\
&& - \ {1\ov \pi} 
\int_{U_H}^{U_0} dU {U^2 \ov \sqrt{U^4-r_0^2 U^2 -\m^4}} \ ,
\label{jewhconf}
\ea
where $U_H$ is given by \eqn{jsab}. 
The behaviour is Coulombic for small separations and
confining for $U_0\to U_H$, since we then have
\be
L\ \simeq \  {R^2 \ov \sqrt{2} } (r_0^4 + 4 \m^4)^{-1/4} 
\ln\left({U_H\ov U_0-U_H }\right)\ ,
\label{jsdconf}
\ee
and
\ba
E_{q{\bar q}} & \simeq & {U_H^2 \ov 2 \sqrt{2} \pi} (r_0^4 + 4
\m^4)^{-1/4} 
\ln\left({U_H\ov U_0-U_H}\right) \nonumber \\
 & \simeq & \frac{U_H^2}{2\pi R^2} L \ ,\qq {\rm for}\quad  
L\gg R^2 (r_0^4+4 \m^4)^{-1/4}  \ .
\label{j1sd1conf}
\ea
As in section 4.1 we found a transition from confinement to
the Coulomb law, but the force between the external charges is a
smooth function of $L$. Also in this case
the comments made in the last paragraph of section 4.1 apply.


\section{Concluding remarks and some open problems}

We have studied Wilson loops and computed the associated 
heavy quark--antiquark potential within the AdS/CFT
correspondence and in the supergravity approximation.

We have studied four-dimensional 
supersymmetric gauge theories at zero temperature with broken gauge symmetry, 
using multicentre D3-brane solutions. In this case we found that there 
is a complete screening of charges that we attributed to the existence of a 
mass gap in the gauge theory. It is very important to understand such 
a mass gap from a field theoretical point of view. 
A first difficulty is that 
the mass gap, and in fact all interesting phenomena, occur at a 
scale $\sim {r_0\ov R^2}$, whereas the vev masses $\sim r_0$ are much larger.
This has a similarity with the fact that two different energy scales occur 
in the AdS/CFT correspondence \cite{PP}.
It is also important to understand to what extent our results are 
a feature of the continuum nature of the brane distribution.

Gauge theories at finite temperature were also studied using 
rotating D3-brane solutions.  
In this case we found two distinct classes of potentials. One is similar
to that obtained by using non-extremal D3-branes with
zero rotation and the other to one that interpolates between a Coulombic
and a confining potential. In the latter case, depending on the 
ratio of the extremality to the rotation parameter, the interpolation 
is smooth or the transition between 
the two regions is reminiscent of phase transitions in statistical
systems. The potential contains three distinct branches,  
one of which violates the concavity condition that states that the force 
between a quark and an antiquark is a monotonously decreasing function of 
their separation. However, the physical path, similarly to
the physical isotherm 
in statistical systems, connects the Coulombic and the confining regions 
of the potential through the coexistence
point, where the potential is smooth, but the force is
discontinuous. It would be extremely interesting to understand this
discontinuity from a field theory point of view.
Using the same rotating brane solution, non-supersymmetric
three-dimensional gauge theories at zero temperature were also studied
by computing the heavy quark--antiquark potentials from 
the associated space--space Wilson loops, yielding similar results.

There are several straightforward, but quite important,
generalizations of our work.
The first is to study the potentials arising from the most general rotating 
D3-brane solution \cite{cvsen,KLT,RS} (eq. (30) in \cite{RS}), which 
contains three rotational parameters. It can be easily seen that there
exist trajectories with all the angles corresponding to $S^5$ fixed.
Second, it will be interesting to compute the heavy quark--antiquark 
potential from the space--space Wilson loop obtained by 
using the rotating D4-brane solution of \cite{CRST}.  This is appropriate
for studying, in the present context, four-dimensional non-supersymmetric
gauge theories at zero temperature.
Finally, we mention that a concavity condition for Wilson loops, 
generalizing \eqn{po2},
has been derived \cite{dorn} for the cases in which 
there is a relative orientation between the quark and 
antiquark with respect to the sphere coordinates.
The prototype computation of such potentials was done,
for the case of the $AdS_5 \times S^5$ background, in \cite{malda},  
and it indeed obeys the condition of \cite{dorn}.
In that respect it is important to know if this generalized 
concavity condition is obeyed 
by the similar potentials one would obtain using rotating D3-brane metrics.

Another important line of research concerns the inclusion 
of corrections to 
the leading order supergravity approximation.
It would be interesting to include $\alpha'$-corrections
to the supergravity backgrounds and to study how they affect our solution.
Note that such corrections are always present for the $N=4$ theory
on the Coulomb branch, since the Weyl tensor does not vanish and
the Gross--Witten term in the supergravity action is non-zero. Only
at the origin of the Coulomb branch is the Weyl tensor exactly zero
and $AdS_5 \times S^5$ an exact background of type IIB string theory.
Although a complete formulation of string theory in the background
of Ramond fields is still missing, it would also be interesting to
calculate corrections due to worldsheet fluctuations. This could
be done, for instance, using a Green--Schwarz formulation and expanding
around the classical configuration. Calculations of this kind
have recently been performed in \cite{fluct}.


\appendix
\section{D3-branes distributed on a ring}

\setcounter{equation}{0}
\renewcommand{\theequation}{\thesection.\arabic{equation}}

In this appendix we compute the quark--antiquark potential for 
$N$ branes distributed, uniformly in the angular direction,
around the circumference of a ring of radius $r_0$ in the $x_5$--$x_6$ 
plane \cite{Sfe1}.
Their centres are given by 
\ba
&& {\bf x}_{i} = (0,0,0,0,r_{0} \cos\phi_i, r_{0} \sin\phi_i )\ ,
\nonumber \\
&& \phi_i = {2\pi i\ov N}\ , \quad  i=0,1,\dots, N \ .
\label{viw1}
\ea
In the near--horizon limit the harmonic function is given by
\ba
&&H = \frac{R^4}{{\alpha'}^2} \frac{U^2 + r_0^2}{[(U^2+r_0^2)^2 
- 4 r_0^2 u^2]^{3/2}} \ \S_N(x,\psi)\ ,
\nonumber \\
&& U^2 = U_1^2 + \ldots + U_6^2 ~,~~~~ u^2 = U_5^2 + U_6^2 ~.
\label{hring}
\ea
where
\be
\S_N(x,\psi) \equiv 
{\sinh(Nx) \ov \cosh(Nx) - \cos(N\psi)}
+N {\Big( (U^2+r_0^2)^2 - 4 r_0^2 u^2\Big)^{1/2}\ov U^2+r_0^2}\ 
{\cosh N x \cos N\psi -1 \ov (\cosh Nx -\cos N\psi )^2} \ .
\label{jhlk}
\ee
The variable $x$ appearing in \eqn{jhlk} is defined as 
\be
e^x \equiv {U^2 + r_0^2 \ov 2 r_0 u} 
+ \sqrt{\left(U^2 + r_0^2 \ov 2 r_0 u\right)^2-1}\ ,
\label{exx}
\ee
whereas $\psi$ is the angular variable in the $x_5$--$x_6$ plane.
In the limit of a continuous distribution of branes where $N\gg 1$, we may 
set $\S_N$ equal to 1. This is a very good approximation unless we 
approach the ring at $U=u=r_0$ down to distances such that 
$U/r_0-1={\cal O}({1\ov N})$ or smaller.

We studied two kinds of trajectories, one where the trajectory is orthogonal 
to the plane in which the ring is lying, i.e. $u = 0$, and one where the 
trajectory runs in a radial 
direction in the plane of the ring towards the centre
of the ring, i.e. $U = u$.
In the first case we find (without setting in \eqn{hring} $\S_N=1$)
for $f(U)$ and $g(U)$ the same functions as in \eqn{rtw1}.
Hence, the corresponding quark--antiquark potential is the same as
the one computed in section 3.2.2.
In the second case we consider only the continuum limit, where $\S_N=1$.
We have 
\be
f(U) = \frac{(U^2 - r_0^2)^3}{(U^2 + r_0^2)}\ , \qq g(U)=1
\ee
and $U_{\rm min}=r_0$. After we change variables as $\r=U^2$,  
we obtain for the length and the energy the integrals 
\be
L = R^2 (U_0^2 - r_0^2)^{3/2} \int_{U_0^2}^\infty \frac{d\rho\ (\rho + r_0^2)}
{\sqrt{\rho (\rho - r_0^2)^3 [(\rho-r_0^2)^3(U_0^2+r_0^2) -
(\rho+r_0^2)(U_0^2-r_0^2)^3]}}
\label{jkwd}
\ee
and
\be
E_{q{\bar q}} = 
\frac{1}{2 \pi} 
\int_{U_0^2}^\infty d\rho \left\{ \frac{\sqrt{U_0^2 + r_0^2}\ 
(\r- r_0^2)^{3/2}} 
{\sqrt{\rho [(\rho-r_0^2)^3(U_0^2+r_0^2) -(\rho+r_0^2)(U_0^2-r_0^2)^3]}} - \frac{1}{\sqrt{\rho}} 
\right\} - \frac{U_0 - r_0}{\pi}  \ .
\ee
For $U_0 \gg r_0$ the potential
is Coulombic and has the same form as in \eqn{qqD3}. 
In contrast to the previous trajectory, we do not find
confinement for $U_0 \simeq r_0$. 
We approximately evaluated the behaviour for $U_0 \simeq r_0$.
The result is
\be 
L\ \simeq \ {2\sqrt{\pi} \G(2/3)\ov \G(1/6)} {R^2\ov \sqrt{r_0 (U_0-r_0)}} 
\label{ja12}
\ee
and 
\be
E_{q{\bar q}}\ \simeq \ - {\G(2/3)\ov \sqrt{\pi} \G(1/6)} (U_0-r_0)\ .
\label{jehf1}
\ee
Therefore
\be
E_{q{\bar q}}\ \simeq \ -  4\sqrt{\pi} \left({\G(2/3)\ov \G(1/6)}\right)^3
{R^4\ov r_0 L^2}\ .
\label{lsq}
\ee
This can be interpreted as a screening of the $L^{-1}$ Coulombic potential 
to an $L^{-2}$ potential at large separation. 
The latter behaviour is characteristic
of a quark--antiquark potential of a gauge theory living on D4-branes 
or on D3-branes 
smeared along a transverse direction. Indeed, for large $L$, when 
the corresponding trajectory approaches the ring circumference, it can be
shown \cite{Sfe1} that the supergravity solution becomes that of a 
D3-brane with a transverse direction smeared out.
One finds that $f(U)=4 r_0 U^3$ and $g(U)=1$. Then applying \eqn{le1} and
\eqn{en1} we confirm \eqn{ja12}--\eqn{lsq}.

Finally, let us mention that there is no known finite--temperature 
supergravity 
solution corresponding to the ring geometry for D3-branes. Owing to the
no-hair theorem, the rotating D3-brane solutions are unique; in the 
supersymmetric limit, they give rise to a uniform distribution of D3-branes,
in the case of one angular momentum, over a disc. 
The ring geometry arises naturally in the supersymmetric limit of rotating
NS5- and D5- branes \cite{sfe2}. Then, the background  
corresponds to the exact conformal 
field theory coset model $SL(2,\IR)/U(1) \times SU(2)/U(1)$ \cite{Sfe1,sfe2}.

\bigskip\bigskip

\centerline{\bf Acknowledgement }

We would like to thank C. Bachas for a discussion. 
On the day we submitted our paper in hep-th we received \cite{FGPW} 
which has some overlap with sections 2 and 3.2 of our paper 
concerning complete screening in Wilson loops and mass gaps in the 
spectrum of gauge invariant operators in backgrounds corresponding 
to continuous distributions of D3-branes.


\begin{thebibliography}{3}


\bibitem{Maldacena}
J.~Maldacena,
Adv. Theor. Math. Phys. {\bf 2} (1998) 231,
hep-th/9711200.

\bibitem{Gubser}
S.S.~Gubser, I.R.~Klebanov and A.M.~Polyakov,
Phys. Lett. {\bf B428} (1998) 105, hep-th/9802109.

\bibitem{Witten}
E.~Witten,
Adv. Theor. Math. Phys. {\bf 2} (1998) 253,
hep-th/9802150.


\bibitem{cvsen}
M. Cvetic and D. Youm, Nucl. Phys. {\bf B477} (1996) 449, hep-th/9605051.


\bibitem{russo}
J.G.~Russo,
Nucl. Phys. {\bf B543} (1999) 183,
hep-th/9808117.


\bibitem{KLT}
{P. Kraus, F. Larsen and S.P. Trivedi, 
JHEP {\bf 03} (1999) 003, hep-th/9811120.}

\bibitem{RS}
{J.G. Russo and K. Sfetsos, 
Adv. Theor. Math. Phys. {\bf 3} (1999) 131, hep-th/9901056.}


\bibitem{Sfe1}
K.~Sfetsos,
JHEP {\bf 01} (1999) 015, hep-th/9811167.

\bibitem{rey}
S.~Rey and J.~Yee,
{\it Macroscopic strings as heavy quarks in large N
gauge theory and Anti-de-Sitter supergravity}, hep-th/9803001. 

\bibitem{malda} 
J.~Maldacena,
Phys. Rev. Lett. {\bf 80} (1998) 4859,
hep-th/9803002.

\bibitem{Bachas}
C.~Bachas,
Phys. Rev. {\bf D33} (1986) 2723.

\bibitem{grosse}
B.~Baumgartner, H.~Grosse and A.~Martin,
Nucl. Phys. {\bf B254} (1985) 528.



\bibitem{Kinar}
Y.~Kinar, E.~Schreiber and J.~Sonnenschein,
{\it Q anti-Q potential from strings in curved space-time: Classical results},
hep-th/9811192.


\bibitem{wilfinTRey}
S.~Rey, S.~Theisen and J.~Yee,
Nucl. Phys. {\bf B527} (1998) 171
hep-th/9803135.


\bibitem{b1} 
A.~Brandhuber, N.~Itzhaki, J.~Sonnenschein and S.~Yankielowicz,
Phys. Lett. {\bf B434} (1998) 36,
hep-th/9803137.



\bibitem{tipologio}
I.S.~Gradshteyn and I.M.~Ryzhik, 
{\it Table of integrals, series and products}, fifth edition 
(Academic Press, New York, 1994).

\bibitem{BF}
P.~Byrd and M.~Friedman, {\it Handbook of Elliptic Integrals for
Engineers and Physicists}, second edition, (Springer Verlag,
Heidelberg, 1971).


\bibitem{KS} A. Kehagias and K. Sfetsos, Phys. Lett. {\bf B454} (1999)
270,
hep-th/9902125. 


\bibitem{Witten2}
E.~Witten,
Adv. Theor. Math. Phys. {\bf 2} (1998) 505,
hep-th/9803131.


\bibitem{tipolo1} M.~Abramowitz and I.A.~Stegun,
{\it Handbook of mathematical functions},
(Dover Publications, New York, 1964).


\bibitem{svet}
B.~Svetitsky, Phys. Rep. {\bf 132} (1986) 1.

\bibitem{minahan}
J.A. Minahan and N.P. Warner, JHEP {\bf 06} (1998) 005, hep-th/9805104.


\bibitem{Callen}
H.B. Callen, {\it Thermodynamics and introduction to thermostatistics}, 
2nd edition, (John Wiley \& Sons, New York, 1985).


\bibitem{COOT98}
C. Cs\'aki, H. Ooguri, Y. Oz and J. Terning, 
JHEP {\bf 01} (1999) 017,
hep-th/9806021;\hfill\break
R.~de Mello Koch, A.~Jevicki, M.~Mihailescu and J. Nunes, 
{Phys. Rev.} {\bf D58} (1998) 105009,
hep-th/9806125;\hfill\break
M. Zyskin, 
Phys. Lett. {\bf B439} (1998) 373, hep-th/9806128.



\bibitem{csaki} C. Cs\' aki, Y. Oz, J.G. Russo and J. Terning, 
Phys. Rev. {\bf D59} (1999) 065008,
hep-th/9810186.




\bibitem{b2}
A. Brandhuber, N. Itzhaki, J. Sonnenschein and S. Yankielowicz,
JHEP {\bf 06} (1998) 001,
hep-th/9803263.

\bibitem{PP} A.W. Peet and J. Polchinski, Phys. Rev. {\bf D59} (1999) 065011.

\bibitem{CRST}
C. Csaki, J.G. Russo, K. Sfetsos and J. Terning, 
Phys. Rev. {\bf D60} (1999) 044001, hep-th/9902067.

\bibitem{dorn}
H.~Dorn and V.D.~Pershin,
Phys. Lett. {\bf B461} (1999) 338, hep-th/9906073.


\bibitem{fluct}
J.~Greensite and P.~Olesen,
JHEP {\bf 04} (1999) 001, hep-th/9901057;\hfill\break
S.~Forste, D.~Ghoshal and S.~Theisen,
JHEP {\bf 08} (1999) 013, hep-th/9903042;\hfill\break
S.~Naik,
Phys. Lett. {\bf B464} (1999) 73, hep-th/9904147.


\bibitem{sfe2}
K. Sfetsos, {\it Rotating NS5-brane solution and its exact string theoretical 
description}, 
Proceedings of the {\it 32nd International Symposium Ahrenshoop on 
the Theory of Elementary Particles},
Buckow, Germany, 1-5 September 1998, hep-th/9903201.

\bibitem{FGPW} D.Z. Freedman, S.S. Gubser, K. Pilch and N.P. Warner,
{\it Continuous distributions of D3-branes and gauged supergravity},
hep-th/9906194.


\end{thebibliography}
\end{document}